\documentclass{ws-jai}
\usepackage[flushleft]{threeparttable}

\usepackage{natbib}
\defcitealias{itu_p452_16}{ITU-R~Rec.~P.452, 2015}
\defcitealias{itu_p1144_9}{ITU-R~Rec.~P.1144, 2017}
\defcitealias{itu_ra769_2}{ITU-R~Rec.~RA.769, 2003a}
\defcitealias{itu_ra1631_0}{ITU-R~Rec.~RA.1631, 2003b}
\defcitealias{itu_f1766_0}{ITU-R~Rec.~F.1766, 2006}
\defcitealias{itu_ra2332}{ITU-R~Rep.~RA.2332, 2011}
\defcitealias{cispr11}{EN\,550011 (CISPR-11, 2015)}

\begin{document}

\catchline{}{}{}{}{} 

\markboth{Winkel \& Jessner}{Compatibility Between Wind Turbines and RAS}

\title{Compatibility Between Wind Turbines and the Radio Astronomy Service}

\author{Benjamin Winkel$^{1}$ and Axel Jessner}

\address{
Max-Planck-Institut f\"{u}r Radioastronomie, Auf dem H\"{u}gel 69, 53121 Bonn, Germany
}

\maketitle

\corres{$^{1}$\texttt{bwinkel@mpifr.de}}

\begin{history}
\received{(to be inserted by publisher)};
\revised{(to be inserted by publisher)};
\accepted{(to be inserted by publisher)};
\end{history}

\begin{abstract}
Modern radio astronomical facilities are able to detect extremely weak electromagnetic signals not only from the universe but also from man-made radio frequency interference of various origins. These range from wanted signals to unwanted out-of-band emission of radio services and applications to electromagnetic interference produced by all kinds of electronic and electric devices. Energy harvesting wind turbines are not only equipped with electric power conversion hardware but also copious amounts of electronics to control and monitor the turbines. A wind turbine in the vicinity of a radio telescope could therefore lead to harmful interference, corrupting the measured astronomical data. Many observatories seek to coordinate placement of new wind farms with wind turbine manufacturers and operators, as well as with the local planning authorities, to avoid such a situation. In our study, we provide examples as well as guidelines for the determination of the separation distances between wind turbines and radio observatories, to enable a benign co-existence for both.

The proposed calculations entail three basic steps. At first, the anticipated maximum emitted power level based on the European \citetalias{cispr11} standard, which applies to industrial devices, is determined. Then secondly, the propagation loss along the path to the radio receiver is computed via a model provided by the international telecommunication union. Finally, the received power is compared to the permitted power limit that pertains in the protected radio astronomical observing band under consideration. This procedure may be carried out for each location around a telescope site, in order to obtain a map of potentially problematic wind turbine positions.
\end{abstract}

\keywords{methods: statistical, telescopes, site testing}

\section{Introduction}\label{sec:intro}
\noindent Harvesting wind energy is one of the few sustainable ways to generate energy with a low CO$_2$ footprint. Considering the challenges and threats imposed by the global climate change, all efforts to utilize renewable energy sources deserve our support. Modern radio astronomy is however extremely sensitive to human-made electromagnetic emission and in the worst case scientists could lose a unique window into the universe if interference swamps the weak signals from space. Wind turbines (WT), as all other industrial devices in the vicinity of a radio telescope, have a potential of interference for astronomical observations. Therefore, coordination is needed to guarantee a beneficial coexistence between radio astronomy and societies interest in wind power utilization. Given the large area that is suitable for wind energy harvesting in most countries, the small fraction of land, which would have to be kept free around radio observatories seems insignificant at first glance. However, not all regions are equally well suited for a WT farm. Radio observatories are often placed in remote locations where wind farms could otherwise be operated without, e.g., affecting existing settlements. The LOFAR core station in the Netherlands is for example located in an area with very high wind harvesting potential, which can easily lead to a serious conflict of interests.

Radio astronomy is fundamental research, which studies the near and far universe. Many cosmic phenomena produce electromagnetic radiation that can only be detected by large radio antennas even though huge amounts of energy are sometimes emitted as the immense distances to the astronomical objects cause an even stronger attenuation of the signals. Thus, the received power of cosmological objects is often many orders of magnitude weaker than that of artificial sources in the environment of a telescope. One illustrative example was the detection of the MASER emission from a huge cloud of water vapor at a distance of $11\cdot 10^9~\mathrm{lyr} = 10^{23}~\mathrm{km}$. The received signal with a strength of about $2~\mathrm{mJy} = -287~\mathrm{dB}\left[\mathrm{W\,m}^{-2}\,\mathrm{Hz}^{-1}\right]$ required 14 hours of observation of the source MG\,J0414$+$0543 with the Effelsberg 100-m radio telescope \citep{impellizzeri08}. Because of the distance the signal has been attenuated by about 556~dB, which allows us to estimate the power of the source to be $305~\mathrm{dB\left[W\right]}$, or 10,000 times the total luminosity of the Sun.

Radio astronomical observatories utilize large dishes or arrays with sometimes hundreds or thousands of smaller antennas, very often equipped with cryogenically cooled low-noise amplifiers. In combination with ultra-stable electronics, which allows us to integrate the incoming signal for several hours in order to decrease the effective noise level, this leads to the enormous sensitivities necessary for the detection of the weak signals from the universe. Furthermore, a large variety of radio frequency interference (RFI) detection and mitigation techniques has been developed in the past two decades, from hardware-based solutions (e.g. superconducting filters) to real-time digital processing and convolutional neural networks \citep[see][and references therein]{fridman01,tuccari04,gary10,offringa12a,offringa12b,akaret17}. Yet, even today the most effective interference mitigation is to ensure that harmful artificial signals are prevented from entering the telescope system in the first place. This is the aim of spectrum management for radio astronomy \citep[see e.g.][]{pankonin81,craf05,driel09,itu_handbook}.

The International Telecommunication Union\footnote{\url{https://www.itu.int/}} (ITU), in particular its radio-communication sector (ITU-R), acknowledged the importance of the radio astronomy service (RAS) already many decades ago in 1959 \citep{craf05}. Protection criteria have been formulated to ensure that some of the frequency bands that are of highest importance to radio astronomy are kept free from RFI. As an example, one of the most important spectral lines for radio astronomy is the 21-cm transition of neutral atomic hydrogen (\textsc{Hi}) having a rest frequency of 1420.4 MHz. To guarantee interference-free measurements of \textsc{Hi}, no man-made emission is permitted in the frequency range between 1400 and 1427~MHz. This and other rules are part of the radio regulations by the ITU-R. Furthermore, a methodology to calculate limits on received interference is described in the recommendation \citetalias{itu_ra769_2}, which is intended to guarantee a certain minimum quality of the recorded data for the astronomers. The governments and administrations of all nations that are represented in the ITU-R have agreed to implement the rules and procedures decided by ITU-R bodies into their national law.

The paper is organized as follows. In Sec.~\ref{sec:compatstudy} we explain how compatibility studies between a transmitting radio service and an RAS station could be performed and which information is necessary for such calculations. The proposed methodology is then applied to the so-called ``flat Earth'' case, where terrain heights are neglected (Sec.~\ref{sec:generic}). This is followed by a case study in Sec.~\ref{sec:casestudy}, where we derive exclusion zones around an existing radio telescope, here the Effelsberg 100-m dish, which is situated in a valley within the Eifel mountains in western Germany, which will serve as an example. A summary is provided in Sec.~\ref{sec:summary}.

\section{Compatibility study methodology}\label{sec:compatstudy}

To study the compatibility between two radio services, it is necessary to calculate the fraction of electromagnetic emissions emanating from a transmitter (\textit{interferer}) that will find their way into the receiving system (\textit{victim}). The received power or the (spectral) power flux density can then be compared to the maximal acceptable level of interference that will still allow proper operation of the victim service \citep{jessner13}. Not all of the power invested into the transmission of a signal can be transformed into the desired radiation, some of it will be simply converted into heat or it may be radiated at frequencies outside of the allocated band. Such ``unwanted'' emissions have to be considered when the victim operates in one of the adjacent frequency bands. The radiated power will also be direction-dependent, based on the characteristics of the antenna in use. Propagation effects attenuate the signal along the path between the transmitter and receiver. The receiving antenna also has a directivity --- providing additional gain if the antenna is pointing towards the transmitter, or attenuation if it isn't, --- which further modifies the link budget.

\subsection{Allowed emissions and RAS protection criteria}\label{subsec:protection}

\begin{wstable}[!t]
\caption{RAS protection limits according to \citetalias{itu_ra769_2} (\textit{excerpt}). Different bandwidths, $\Delta f$, antenna temperatures, $T_\mathrm{A}$, and receiver noise temperatures, $T_\mathrm{Rx}$, apply for each frequency. Based on these numbers, the RMS noise, $T_\mathrm{rms}$, is calculated and the thresholds for the power, $P_\mathrm{lim}$, the power flux density, $S_\mathrm{lim}$, and electrical field strength, $E_\mathrm{lim}$, are derived.}
\begin{tabular}{@{}cccccccc@{}} \toprule
Frequency & $\Delta f$& $T_\mathrm{A}$& $T_\mathrm{Rx}$&
$T_\mathrm{rms}$& $P_\mathrm{lim}$& $S_\mathrm{lim}$& $E_\mathrm{lim}$\\
MHz & MHz& K& K& mK& $\mathrm{dB}_\mathrm{W}$& $\mathrm{dB}_\mathrm{W/m^2}$& $\mathrm{dB}_{\mu \mathrm{V} / \mathrm{m}}$\\ \colrule
\hphantom{0}325  & \hphantom{0}7  & 40 & 60 & 0.870 & $-$201.0 & $-$189.3 & $-$43.5\\
\hphantom{0}408  & \hphantom{0}4  & 25 & 60 & 0.962 & $-$202.9 & $-$189.2 & $-$43.4\\
\hphantom{0}611  & \hphantom{0}6  & 20 & 60 & 0.730 & $-$202.2 & $-$185.0 & $-$39.2\\
1414             & 27             & 12 & 10 & 0.095 & $-$204.5 & $-$180.1 & $-$34.3\\
1665             & 10             & 12 & 10 & 0.156 & $-$206.7 & $-$180.8 & $-$35.0\\ \botrule
\end{tabular}
\label{tab:ras_thresholds}
\end{wstable}

Acceptable interference levels, so-called \textit{limits} or \textit{thresholds}, for RAS stations are given in \citetalias{itu_ra769_2}. With Tab.~\ref{tab:ras_thresholds} we provide an excerpt for the reader's convenience. For a typical receiver operating at a given frequency, the interference power must not exceed 10\% of the RMS noise level that is obtained for a particular (interference-free) bandwidth after a certain observing/integration time \citep[see also][]{jessner13}. Note that the threshold values in \citetalias{itu_ra769_2}, pertain by definition to the combination of a frequency average over the reference bandwidth and a time average over the total integration time.

For spectroscopy observations the bandwidths are of the order of a few kHz --- the spectral channel width --- while in all other cases the bandwidth is set to the size of the protected RAS band, which can range from few MHz at lower frequencies to several GHz at high frequencies. It is common practice by many administrations to use the pre-computed values provided in the tables of \citetalias{itu_ra769_2}, which have been calculated for an integration time of 2000~s with the specific bandwidths for the main observation methods and for each frequency. This is intended to represent a typical case, but longer (see example in the introduction) or shorter times and different bandwidths are also feasible and radio astronomical observations are by no means meant to be restricted to this choice of parameters. \citetalias{itu_ra769_2}, gives details how interference thresholds are calculated for different bandwidths and integration times.

There are a number of frequency bands in which RAS is given explicit protection in the radio regulations of ITU-R, but we will restrict ourselves to consider only the L-Band frequency range 1400$-$1427 MHz as an example in this text.

In terms of electromagnetic interference (EMI) regulations, wind turbines count as industrial devices (Group 1, Class A), which are required to conform to the \citetalias{cispr11} standard: the electrical field strength measured at a distance of 30~m with a quasi-peak (QP) detector having a bandwidth of 120~kHz must not exceed $30~\mathrm{dB}_{\mu \mathrm{V} / \mathrm{m}}$ below 230~MHz or $37~\mathrm{dB}_{\mu \mathrm{V} / \mathrm{m}}$ between 230~MHz and 1~GHz. Limits are not explicitly specified for frequencies above 1 GHz. Furthermore, for other applications above 1~GHz the International Special Committee on Radio Interference (CISPR) typically uses a larger bandwidth (1~MHz) for the measurement channel. We will therefore assume in the following, that the spectral power flux density limits above 1~GHz are the same as the ones between 230~MHz and 1~GHz, which leads to a field strength limit of $27.8~\mathrm{dB}_{\mu \mathrm{V} / \mathrm{m}}$ (per 1-MHz channel). It should also be noted that above 1~GHz CISPR norms tend to refer to Peak or RMS detectors instead of QP.

The electromagnetic power, $P$, that needs to be emitted for the generation of an electrical field strength, $E$, measured at a distance, $d$, is given by
\begin{equation}
P=4 \pi d^2 \frac{E^2}{R_0}\,,
\end{equation}
with $R_0=\sqrt{\frac{\mu_0}{\epsilon_0}}=376.73~\Omega$ being the free space impedance. The electrical field strengths quoted in \citetalias{cispr11} can thus be converted to emitted power levels, which makes a comparison with the \citetalias{itu_ra769_2}, levels easier. In engineering notation\footnote{We note that $E[\mathrm{dB}_{\mu \mathrm{V} / \mathrm{m}}]\equiv10\log(E^2[\mu \mathrm{V} / \mathrm{m}])=20\log(E[\mu \mathrm{V} / \mathrm{m}])$.} this reads
\begin{equation}
P_\mathrm{C11}[\mathrm{dB}_\mathrm{W}] = E_\mathrm{C11}[\mathrm{dB}_{\mu \mathrm{V} / \mathrm{m}}] + 20\log(d_0~[\mathrm{m}]) - 134.8\,.
\end{equation}

The CISPR-11 electrical field limits are defined for a certain measurement bandwidth also specified by CISPR-11. In order to compare them with RAS thresholds from \citetalias{itu_ra769_2}, one will have to calculate the power that is going to be emitted over the appropriate RAS bandwidth. RAS interference limits are defined as averages, while CISPR-11 limits below 1~GHz refer to quasi-peak detections, where the envelope of the received power as a function of time from a chosen detector band is fed to an additional post-detection integrator that has several time constants\footnote{Here we are using the CISPR-11 specification for the appropriate bands C and D (30 MHz -- 1000 MHz).}: a $T_c = 1~\mathrm{ms}$ rise time of the output for any input signal that is greater than the current output level, and a $T_d = 550~\mathrm{ms}$ decay time when the input level is below the output level. This results in a saw-tooth output for time variable inputs which is further smoothed with a ``meter'' time constant $T_m = 100~\mathrm{m}$ \citep{krug04}.

CISPR-11 specifies only QP limits for the relevant bands C and D, corresponding to the peak power for continuous or high repetition rate ($\langle T_\mathrm{rep}\rangle \ll T_d$) which can be about 3~dB higher than the average power, but tend to reflect the averaged power for low duty cycle ($\langle T_\mathrm{rep}\rangle > T_d$) signals \citep{ristau97}. As most interference emissions are by nature highly variable in time and frequency, we assume that the CISPR-11 QP limits may also provide a realistic upper limit for the long ($\approx 2000~s$) time averages specified for radio astronomy protection. The CISPR-11 limits take neither the temporal nor the spectral signature of possible EMI emissions into account. Furthermore, it is not known, whether the electrical and electronic equipment in the WT hub and foot produces a flat spectrum or only a limited number of relatively narrow spectral lines, such as harmonics from oscillators in digital equipment.

We will analyze two example scenarios to illustrate the possible range of compatibility requirements. For \textit{case~1}, we assume that the EMI spectrum is flat and as such every CISPR-11 channel (with 1~MHz channel width) in the RAS band contains the maximal permitted electrical field strength. This corresponds to the highest permissible emission within the full RAS bandwidth. In \textit{case~2}, only one CISPR-11 channel has maximum intensity. Clearly, case~1 is a worst-case scenario and if a WT turns out to be compatible with RAS observations under this assumption, no further action is necessary. For situations where the RAS thresholds are exceeded for case~1, the WT operator may provide further information about the actual spectral properties, which can be taken into account in further calculations.

\begin{figure}[!t]
\begin{center}
\includegraphics[width=0.7\textwidth,viewport=20 20 653 525,clip=]{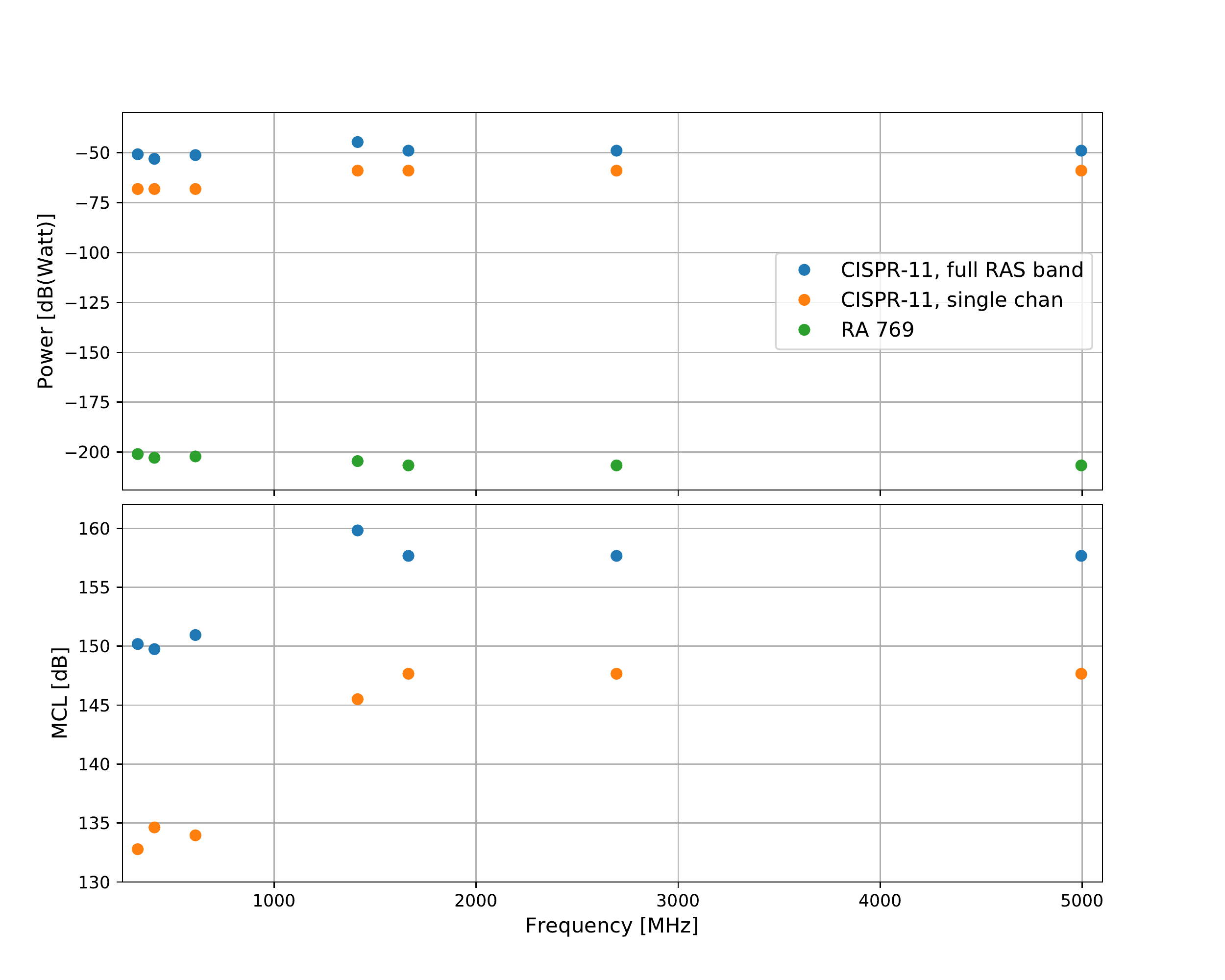}
\end{center}
\caption{\textit{Top panel:} Equivalent \textit{emitted} power levels in RAS bands corresponding to CISPR-11 field strength limits for both scenarios: \textit{full RAS band} (broad-band RFI emission) means the permitted emission levels are reached for each and every CISPR-11 sub-band; \textit{single channel} (single line interference) means the permitted CISPR-11 emission level is reached in only one channel (of width 120~kHz below or 1~MHz above 1~GHz observing frequency), while emissions in other parts of the RAS band are negligible. We also show the limit for received RFI power in the RAS band \citepalias{itu_ra769_2}. \textit{Bottom panel:} The difference between emitted power and permitted received power constitutes the minimal coupling loss (MCL).}%
\label{fig:MCL}%
\end{figure}

In Fig.~\ref{fig:MCL} (top panel), the CISPR-11 limits (converted to power over the full RAS bandwidth) are displayed for a range of protected RAS bands. The RAS protection threshold is also shown in the graph.

\subsection{Minimal coupling loss}\label{subsec:mcl}

The difference between the emitted signals and the RAS threshold (i.e. the gap between the CISPR-11 values and the \citetalias{itu_ra769_2}, thresholds in Fig.~\ref{fig:MCL}, top panel) indicates the amount of additional attenuation that is required to avoid exceeding the ITU-R RAS limits at the telescope. This minimal attenuation is called the minimal coupling loss (MCL),
\begin{equation}
\mathrm{MCL}[\mathrm{dB}] = P_\mathrm{em}[\mathrm{dB}_\mathrm{W}] - P_\mathrm{RA.769}[\mathrm{dB}_\mathrm{W}]\,.
\end{equation}
The main attenuation mechanism is the path propagation loss between the emitter and the receiver, but the antenna gains will also play a role for the coupling loss. However, here we assume a generic omni-directional 0~dBi gain for both patterns. On the one hand, WTs are not antennas by design and to our knowledge no studies exist that examine the effective pattern of radio emission emanating from a WT when radio frequency currents are excited in the structure. For the RAS station, on the other hand, the antenna pattern can be measured and modeled, and there is an ITU-R recommendation \citepalias{itu_ra1631_0}, which provides a simplified description of a ``standard antenna'' for compatibility studies. Nevertheless, the RAS antenna usually points towards an astronomical object of interest located anywhere on the visible sky, but in motion w.r.t. the horizontal coordinate system, while WTs are located close to the local horizon as seen from the telescope. A WT is therefore more or less equally likely positioned anywhere in the antenna pattern, resulting in an \textit{average receiving gain of} 0~dBi. There are however conceivable situations, where a radio telescope observes a source close to the horizon and in a flat countryside it could even happen that the WT is within the main-beam direction. Then, the full forward antenna gain (up to $\sim$90~dBi, depending on dish size and frequency) applies and the detection probability would be substantially increased towards near certainty. Our proposed method for studying more generic scenarios may easily be adapted to more complex cases, but we will omit them here for the sake of clarity and brevity.

In order to assess the impact of operating a wind turbine at a proposed location one needs to compare the MCL with the predicted path propagation loss between the wind turbine and the RAS station. We obtain two different MCL by considering the two scenarios discussed above: the MCL for scenario~1 is by $10\log\left(\Delta f_\mathrm{RAS} / \Delta f_\mathrm{C11}\right)$~dB larger than the corresponding one for scenario~2. Using the reference bandwidths $\Delta f_\mathrm{RAS}$ given in Tab.~\ref{tab:ras_thresholds} of \citetalias{itu_ra769_2}, the difference amounts to $+$14.3~dB for 1420~MHz. Figure ~\ref{fig:MCL} (bottom panel) shows the resulting MCL for both scenarios for a range of RAS frequency bands. In both cases we assume 0 dBi (off-beam) antenna gain for the radio telescope.

\subsection{Path propagation}\label{subsec:pathprop}

For this study, the calculation of path propagation losses between two terminals is based on the method described in \citetalias{itu_p452_16}. It accounts for a variety of propagation/attenuation mechanisms such as
\begin{enumerate}
\item Line-of-sight (free-space) loss including correction terms for multi-path and focusing effects,
\item Diffraction (at terrain features),
\item Tropospheric scatter, and
\item Anomalous propagation (ducting, reflection from elevated atmospheric layers).
\end{enumerate}

While for many other compatibility studies, the effect of clutter has to be incorporated into the calculation, modern WT have heights that significantly exceed typical clutter heights. In our study, we assume that the hub (nacelle) of the wind turbine will be the most important source of radio interference whereas the base and lower parts of the support structure may also radiate, but their radiation will be more strongly absorbed by local clutter and topography. The receiver terminal (the radio telescope) is also higher than surrounding clutter to avoid picking up thermal radiation from objects in the vicinity. We hence assume zero clutter loss for the all following calculations. The scattering at rain drops (so-called hygrometeor scattering) can sometimes play a role, but this effect will also be neglected here. Note, that attenuation by the atmosphere, caused by the oxygen and water content in the lower layers of the atmosphere, is automatically accounted for in the line-of-sight and diffraction terms of the \citetalias{itu_p452_16} propagation algorithm. Its contribution is however negligible at the frequencies under consideration for this work.

A Python implementation of \citetalias{itu_p452_16}, exists, in the form of the \texttt{pycraf} library \citep{winkel18}, which is available as open-source software (GPL-v3) on the Python package distribution server PyPI\footnote{\url{https://pypi.org/project/pycraf/}} (Python Package Index). The software repository is hosted on GitHub,\footnote{\url{https://bwinkel.github.io/pycraf/}} along with detailed documentation and tutorials.

A number of parameters has to be provided in order to calculate the path attenuation loss according to \citetalias{itu_p452_16}, the terrain height profile along the propagation path, the frequency, and the heights of transmitter and receiver above ground being the crucial ones. Furthermore, the so-called time percentage, $p$, for anomalous propagation should be specified (values must be in the range between 0.001\% and 50\%). The path loss resulting from the \citetalias{itu_p452_16}, algorithm has to be understood in statistical terms: the loss value $L(p)$ returned for a given $p$ means that only with a probability of $p$ will the true path loss be higher than $L(p)$. The function $L(p)$ is in fact the inverse of the cumulative distribution function of the loss values. For radio astronomical observations we need to ensure that the RAS thresholds are not exceeded for most of the time, so that only a small fraction of data may be lost. Therefore, we use $p=2\%$ for all subsequent calculations. This choice of percentage is also typical for the regulatory constraints by the ITU on interference probabilities from other services that might affect RAS. Other input parameters include the atmospheric conditions such as temperature, pressure, and humidity, and the average radio-refractive index lapse-rate through the lowest 1 km of the atmosphere, as well as the sea-level surface refractivity. The latter values have relatively small impact on the results (if within reasonable limits), especially at the frequencies used in this study. For details we refer to the \texttt{pycraf} documentation and \citetalias{itu_p452_16}.

\section{Generic results}\label{sec:generic}

\noindent In most cases, a number of existing or prospected WT positions around a specific telescope site will be studied, as the topography around the radio telescope plays an important role in the actual link budget. An initial consideration of the so-called generic case, in which terrain heights are neglected (``flat Earth'' scenario) can however be already quite illustrative. These generic results provide a first estimate of the separation distances that may perhaps be required, and can be a useful guideline for all involved parties: radio astronomers, wind turbine manufacturers/operators, and local planning authorities. We also note, that many compatibility studies submitted to the various working groups of the ITU-R are generic-case, only.

\begin{figure}[!t]
\begin{center}
\includegraphics[width=0.49\textwidth,clip=]{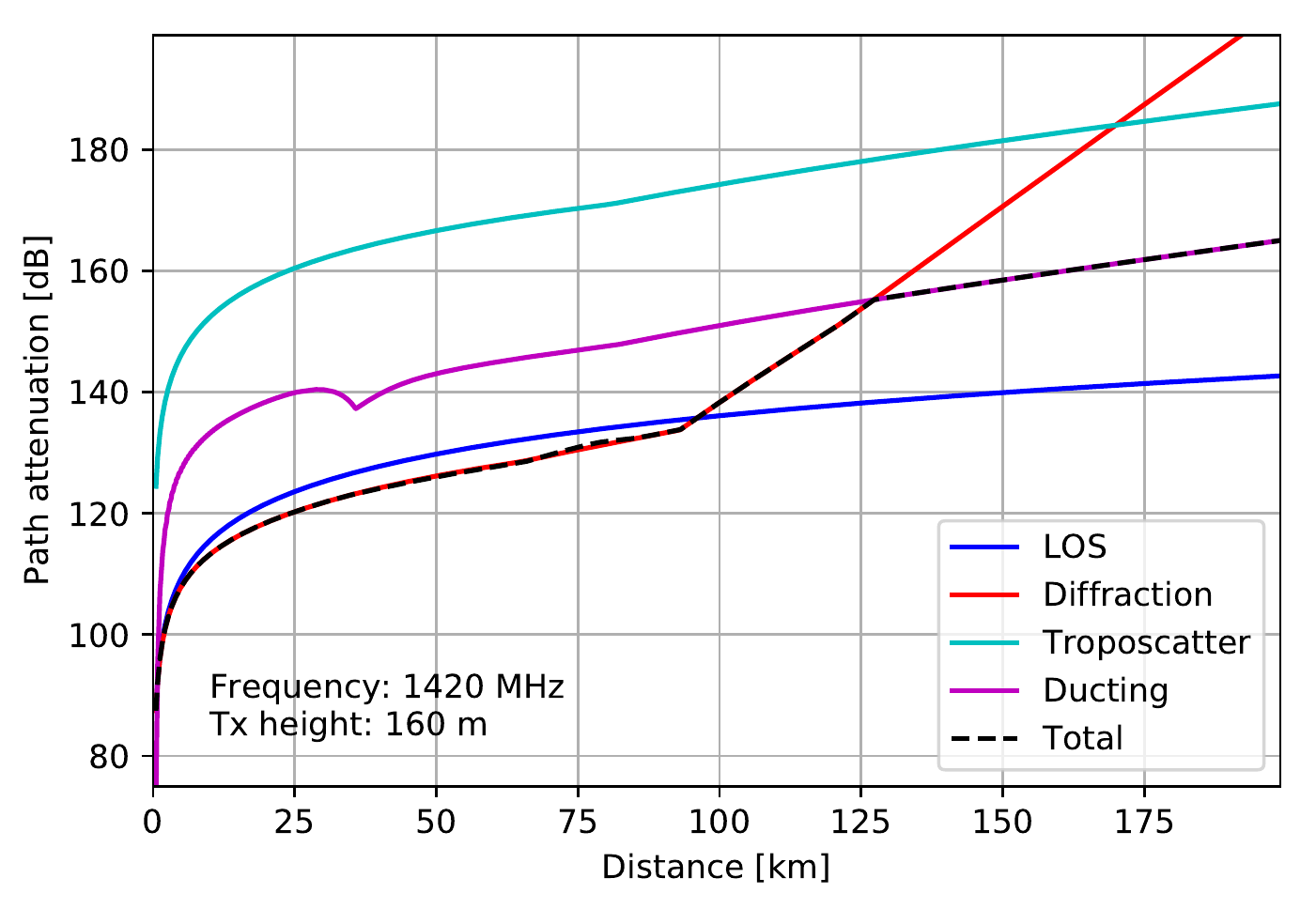}\hfill
\includegraphics[width=0.49\textwidth,clip=]{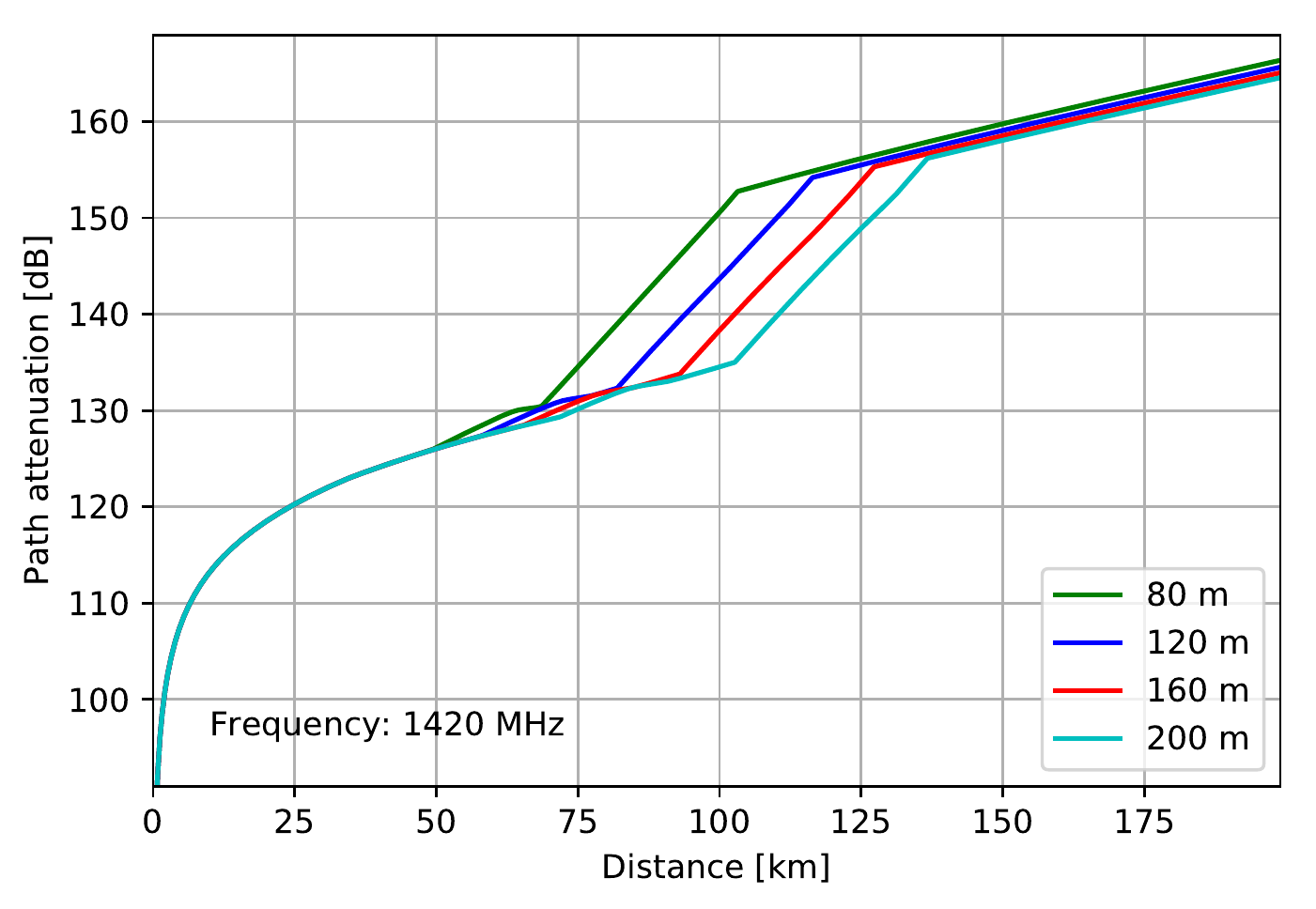}
\end{center}
\caption{\textit{Left panel:} Path propagation loss (dashed black line) as a function of distance for the generic (flat-Earth) scenario at a frequency of 1420~MHz. A transmitter height of 160~m was assumed, and a receiver height of 50~m. The solid lines visualize the various physical effects, e.g. line-of-sight attenuation and diffraction. Note, that the total loss is a non-trivial function of the constituents \citepalias[see][their Sec.~4.6]{itu_p452_16}. \textit{Right panel:} Same as left panel but showing only the total path propagation loss for various transmitter heights.}%
\label{fig:generic_path_atten_components}%
\end{figure}

Figure~\ref{fig:generic_path_atten_components} (left panel) shows the total propagation loss and its constituents over a range of 200~km, calculated for a frequency of 1420~MHz. A receiver height of 50~m and a transmitter height of 160~m have been assumed. The contribution by diffraction increases significantly once the path changes from line-of-sight type to trans-horizon. Subtracting the MCL from the path loss values yields the so-called (link-)margin
\begin{equation}
\mathrm{Margin}[\mathrm{dB}]=L[\mathrm{dB}] - \mathrm{MCL}[\mathrm{dB}]\,.
\end{equation}
If it is zero, then a WT would just be compatible with the RAS power limits at that location. Negative margins indicate a situation where compatibility is compromised.

\begin{figure}[!tp]
\begin{center}
\includegraphics[width=0.49\textwidth,clip=]{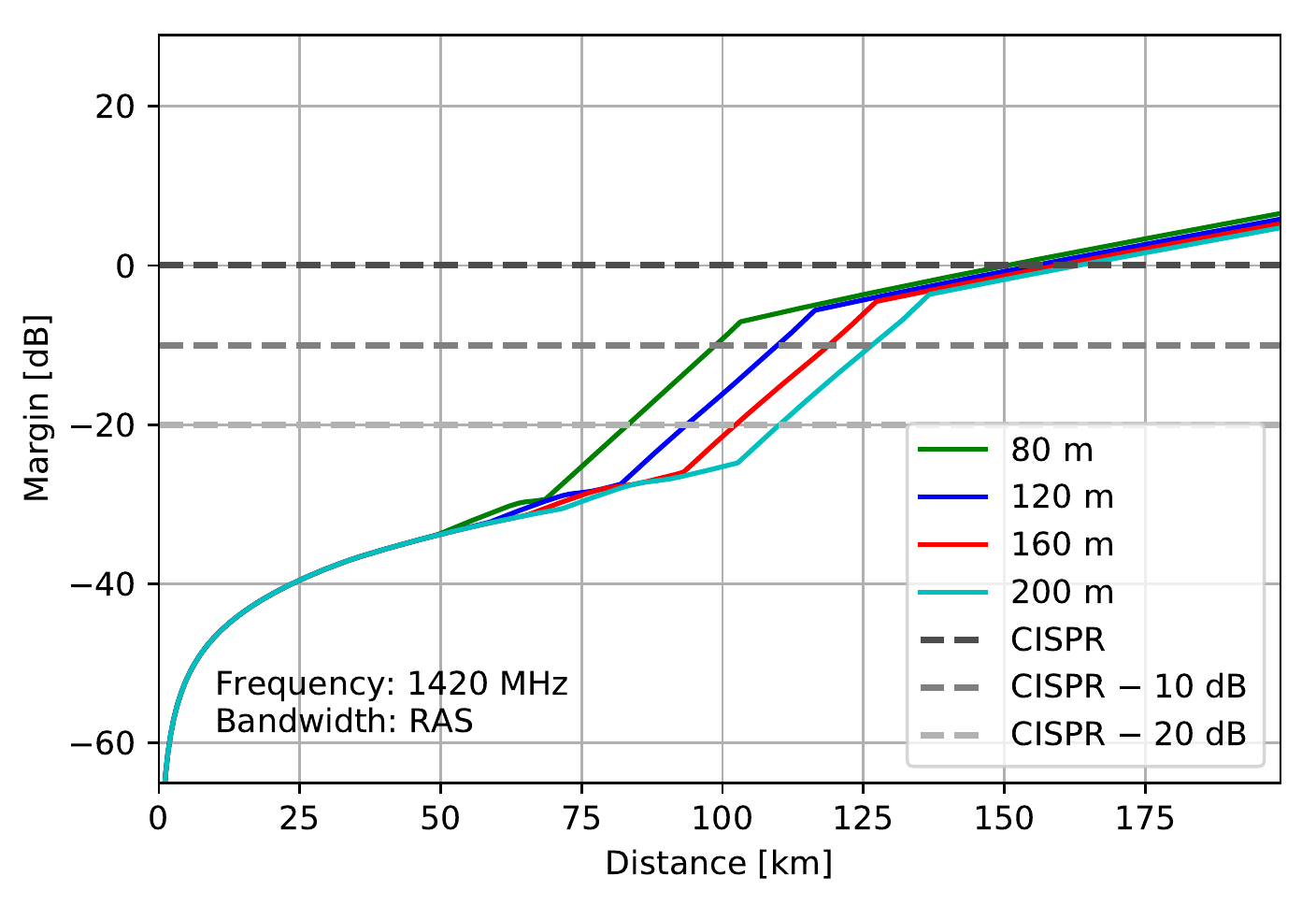}\hfill
\includegraphics[width=0.49\textwidth,clip=]{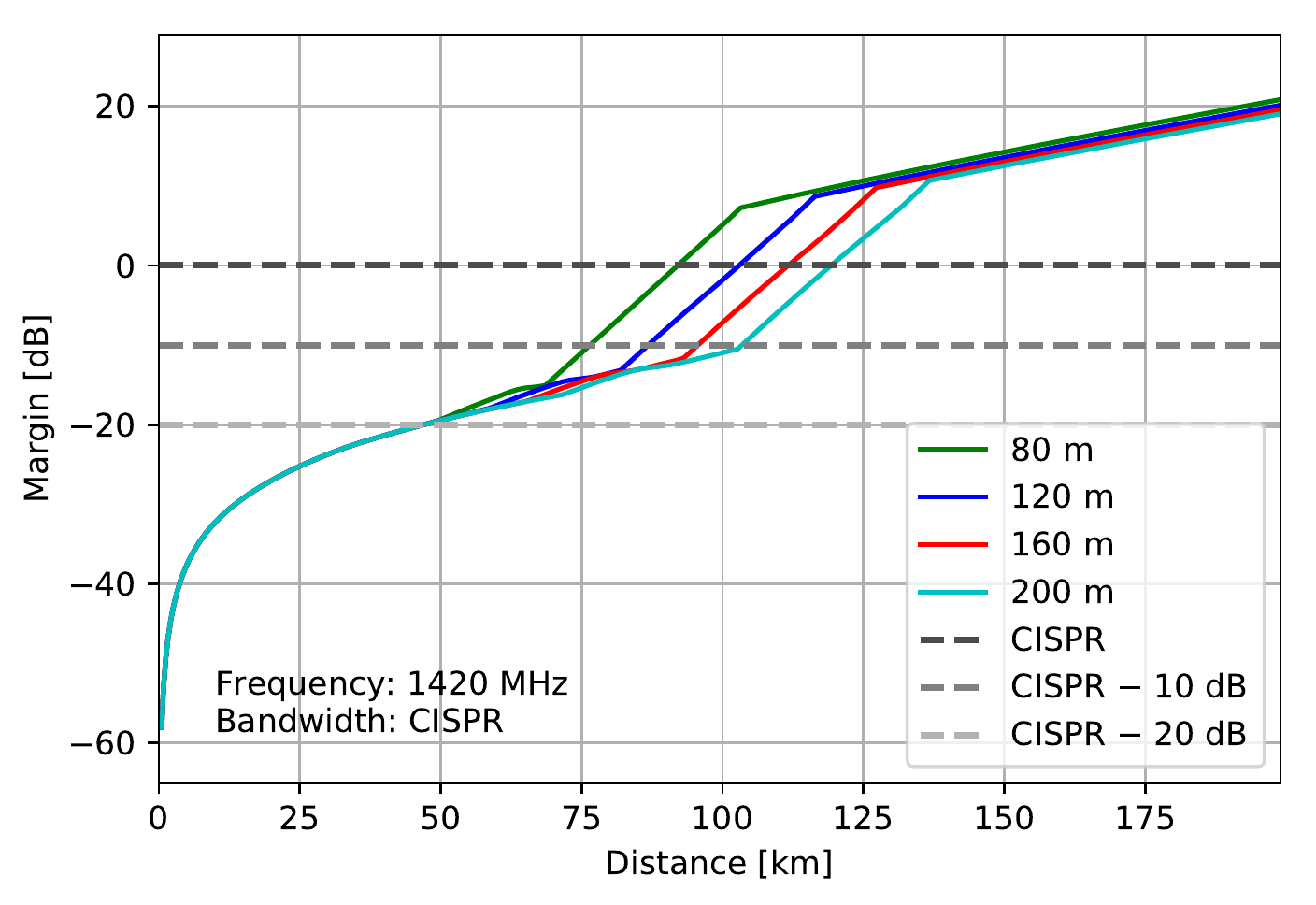}
\end{center}
\caption{Margins for both scenarios and various transmitter heights based on the path propagation loss in Fig.~\ref{fig:generic_path_atten_components}.}%
\label{fig:generic_margin_1420mhz}%
\end{figure}

In Fig.~\ref{fig:generic_margin_1420mhz}, the resulting margins are displayed for both scenarios (CISPR-11 and RAS bandwidths) for various transmitter heights. The dashed black horizontal line marks the zero-margin. Everything below that line could potentially lead to a violation of the thresholds at the radio telescope. The results of several measurement campaigns \citep[e.g.][]{bnetza_report} indicate that the wind turbines may in reality emit 10 to 20~dB less than permitted by CISPR-11. The gray dashed lines would be the separating limits in that case. The number of wind turbines, $N_\mathrm{dev}$, per farm also plays an important role here. For multiple emitters, aggregation effects have to be considered. They lead to an increase of the received power by $\sim$$10\log(N_\mathrm{dev})~[\mathrm{dB}]$ assuming that all devices are subject to the same path propagation loss.

\section{Case studies using topography}\label{sec:casestudy}

The generic analysis discussed in the Sec.~\ref{sec:generic} can only provide a rough estimate of necessary separation distances, because in reality the radio telescopes are not situated in a completely flat environment. Therefore, the topography around a specific site must be taken into account, and that can have a substantial influence on the path attenuation, especially when one considers the diffraction on terrain obstacles.

In this section, we will show how such calculations can be done for a specific observatory site by using the Effelsberg 100-m radio telescope as an example. The 100-m dish is located at the northern edge of the Eifel mountains in the western part of Germany, in a valley near the small town of Bad M\"{u}nstereifel--Effelsberg. The site was chosen to provide a certain level of natural shielding, with the first diffraction edge relatively close to the telescope. However, the telescope is so large that the shielding is not perfectly efficient, especially at frequencies below a few hundred MHz where the diffraction at a single hill top provides only a few Decibels of attenuation. The visual horizon as seen from the dish center is not flat, opening into a valley around azimuths of about $180^\circ$ (i.e. towards the south). In fact, the site was chosen that way, to offer the opportunity to observe the galactic center for up to two hours per day, when it rises only slightly above $10^\circ$ in elevation in that direction.

\begin{figure}[!tp]
\begin{center}
\includegraphics[width=0.8\textwidth,clip=]{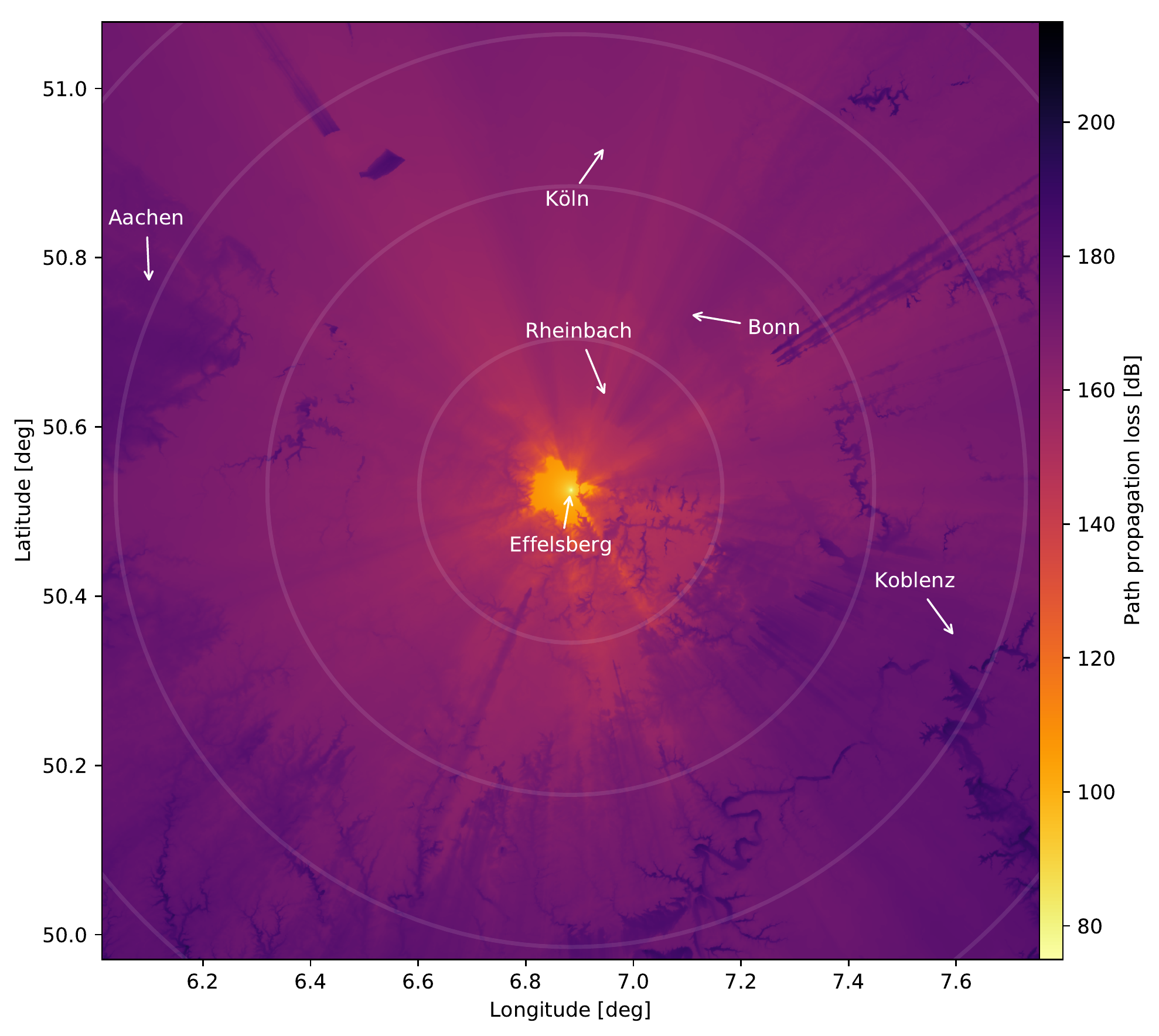}
\end{center}
\caption{Path attenuation map (dB) for the Effelsberg 100-m site using a transmitter height of 160~m and a frequency of 1420~MHz. The white circles indicate distances to the RAS station in steps of 20~km.}%
\label{fig:atten}%
\end{figure}

As for the generic case, the total path propagation loss is calculated according to \citetalias{itu_p452_16}, employing the \texttt{pycraf} software package. \texttt{pycraf} can make use of terrain height data such as provided by the SRTM Space Shuttle Mission \citep{farr07} and incorporate the local topography that way. We discovered by comparing SRTM with other topography data sets that the SRTM data for the chosen RAS site has larger than usual height errors; see Appendix~A. For this particular study we will therefore make use of Light Detection and Ranging (LIDAR) based topography data, which is kindly provided by the German states of North-Rhine-Westphalia\footnote{\url{https://www.opengeodata.nrw.de/produkte/geobasis/dgm/dgm1/}} and Rhineland-Palatinate\footnote{\url{https://lvermgeo.rlp.de/de/geodaten/opendata/}} (Germany). Figure~\ref{fig:atten} displays an example of an attenuation map, which contains the path loss for a transmitter height of 160~m at any position in an area around the radio telescope. The receiver height used in the calculation is again set to 50~m. Because of the specific terrain in the Eifel mountains, the distribution of resulting path attenuations is rather inhomogeneous.

\begin{figure}[!tp]
\begin{center}
\includegraphics[width=0.8\textwidth,clip=]{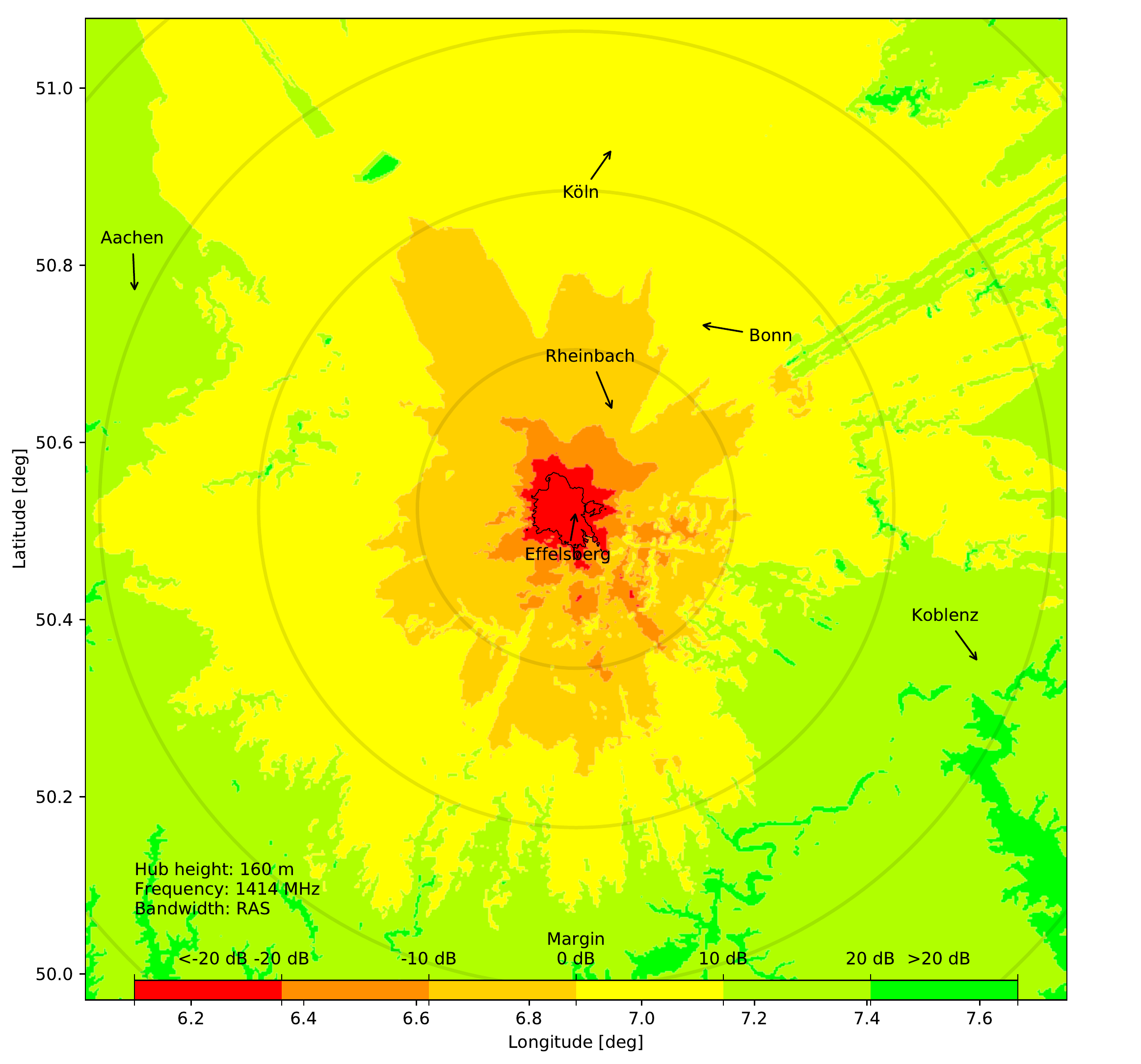}
\end{center}
\caption{Example map of the margin for scenario 1 (RAS bandwidth) at a frequency of 1420~MHz. The transmitter height is 160~m. The black contour marks the (radio-wave) horizon as seen from the center of the telescope dish. The black circles indicate distances to the RAS station in steps of 20~km.}%
\label{fig:margin_1414_mhz_160m_ras}%
\end{figure}

\begin{figure}[!tp]
\begin{center}
\includegraphics[width=0.8\textwidth,clip=]{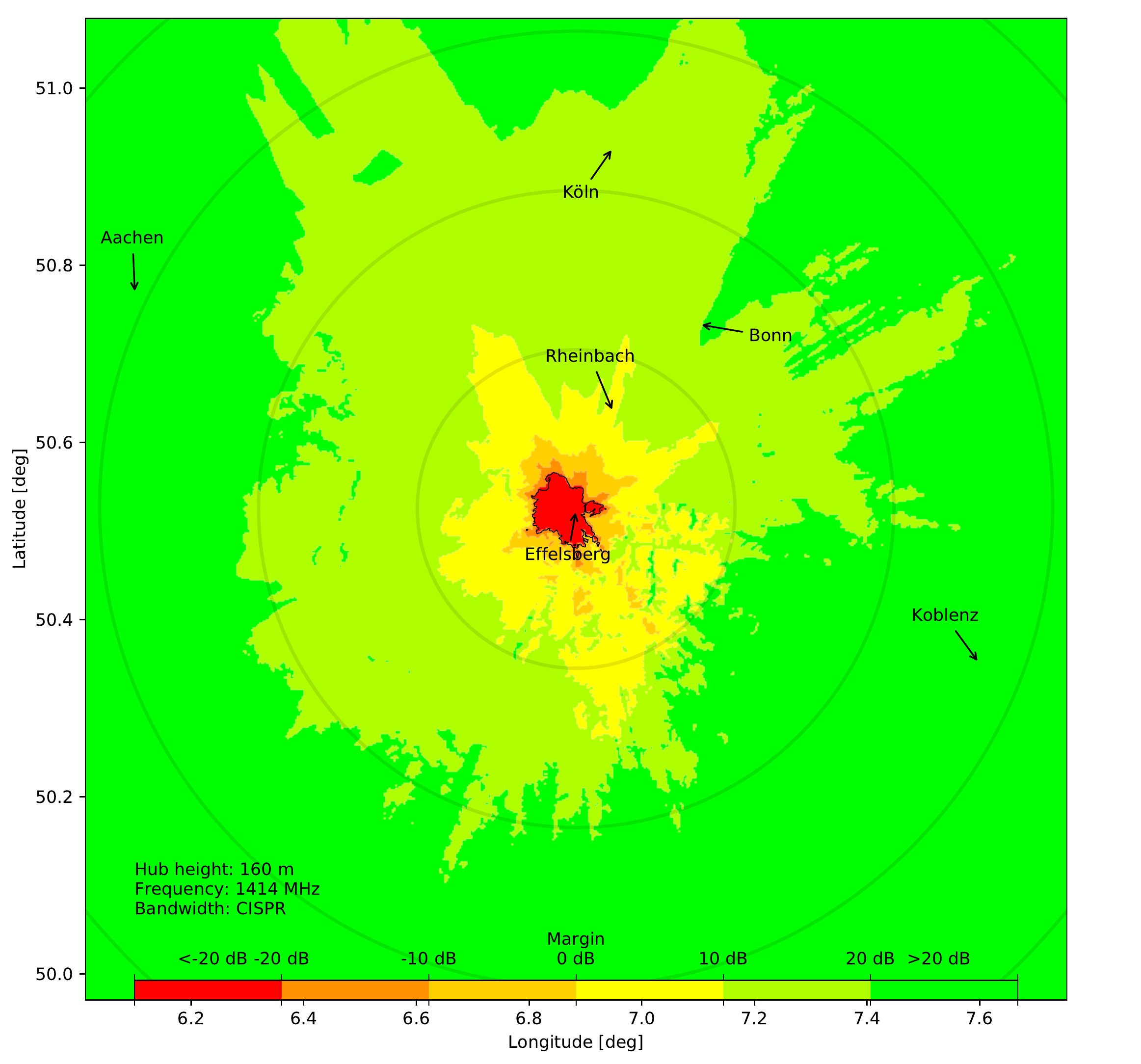}
\end{center}
\caption{As Fig.~\ref{fig:margin_1414_mhz_160m_ras} for scenario 2 (CISPR bandwidth).}%
\label{fig:margin_1414_mhz_160m_cispr}%
\end{figure}

\begin{figure}[!tp]
\begin{center}
\includegraphics[width=0.8\textwidth,clip=]{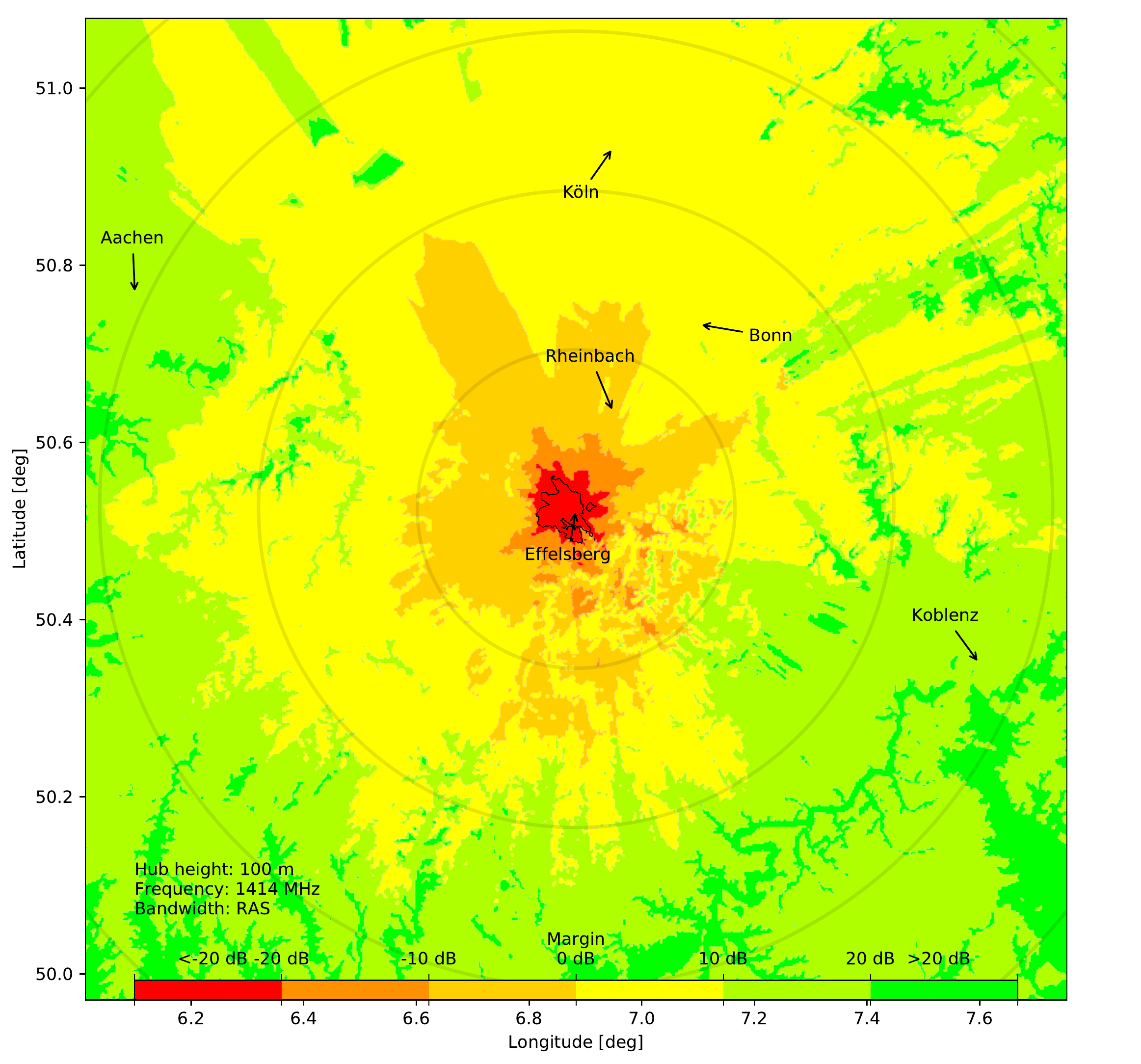}
\end{center}
\caption{As Fig.~\ref{fig:margin_1414_mhz_160m_ras} for a transmitter height of 100~m.}%
\label{fig:margin_1414_mhz_100m_ras}%
\end{figure}

We can calculate the margins for both scenarios using the path attenuation maps (one for each transmitter height). The results are plotted for a transmitter height of 160~m in Figs.~\ref{fig:margin_1414_mhz_160m_ras} and \ref{fig:margin_1414_mhz_160m_cispr} both scenarios), while Fig.~\ref{fig:margin_1414_mhz_100m_ras} contains the margins for a transmitter height of 100~m (scenario~1, only). The critical areas below 0~dB margin have been colored in red and orange, while positive margins are indicated with yellow and green. The (radio-wave) horizon (as seen from the telescope center) is indicated with a black contour for information. We find that the $-$20~dB margin roughly coincides with the horizon for any hub height when emission levels correspond to the scenario~2 CISPR-11 limits.

The number of wind turbines also plays a role as we have already mentioned. When for instance ten wind turbines are clustered in an area with similar path attenuations we would expect to receive about 10~dB more interference power and the 10~dB more stringent contour levels in Figs.~\ref{fig:margin_1414_mhz_160m_ras} to \ref{fig:margin_1414_mhz_100m_ras} would thus define the exclusion zone. If, on the other hand, the wind turbine plant qualifies for emission levels that are 10~dB below CISPR-11, the 10~dB relaxed contour becomes relevant for the definition of the exclusion zone.

The antenna gain pattern of the RAS station causes another effect, that may modify the received power levels. The calculations were done for the 0~dBi case, as explained in Sec.~\ref{subsec:mcl}, because the WT could in general be located at any point in the (forward) pattern and hence both, amplification or attenuation, are possible. If a source is however observed at low elevation close to the horizon, the likelihood that the WT may be in the near side-lobe pattern or even in the main beam of the antenna is not negligible. In that case one possible approach would be the calculation of the so-called path horizon angles for the RAS station, as provided by the \citetalias{itu_p452_16}, algorithm. This is either the elevation angle under which the transmitter appears (for the line-of-sight case), or the angle under which the first diffraction edge is visible (for trans-horizon paths). The effective gain can then be derived for specific telescope pointings \citepalias[a simplified antenna model for RAS telescopes is defined in][]{itu_ra1631_0}. For observatories in deep and narrow valleys in mountainous areas such as the Effelsberg station this approach may however be a severe oversimplification. The 100-m dish is a major structure in the valley, and it is likely that there is some wave coupling between the dish and the surrounding mountains, which are furthermore covered with trees and other clutter sources. Performing a proper 3-D electromagnetic wave simulation to fully explore the overall amplification of the signal in each direction for low elevations is without real alternatives in such cases. The dish can also be partially blocked by local terrain features (i.e. the first Fresnel zone of the path is significantly filled) on low elevations, which is not fully accounted for in the \citetalias{itu_p452_16}m calculations. Likewise, antenna side lobes might be directed to the ground, which increases the system temperature of the RAS receiver, making the observations less sensitive \citepalias[one could therefore argue that the RAS thresholds given by][should be re-calculated with higher system noise parameters in such a case]{itu_ra769_2}.

\section{Summary}\label{sec:summary}

\noindent A method to calculate the potential level of interference that a wind turbine farm could produce in a radio astronomical receiving system has been proposed in this work. It is based on three basic steps: (1) estimating the amount of power that could be produced in the WT (electric and electronic devices in the hub), (2) determining the attenuation between WT and RAS receiver, i.e. the path propagation loss, and (3) comparing the consequently received power with the permitted RAS power limits given in \citetalias{itu_ra769_2}. For the first step, measurements of the emitted power would be needed. This will vary for different types of WT, and a significant effort is required to perform these measurements. We therefore used the maximally permitted power that an industrial device may emit, as given in \citetalias{cispr11} (CISPR-11) in our work.

The required separation distances between a single WT and the RAS terminal are of the order of 150~km if the WT fully exploits the CISPR-11 limits and local terrain effects are neglected (flat earth case). Measurements of real installations indicate that the true emitted power is often at least ten or more Decibels below CISPR-11 levels, in which case the required separation distances shrink to values between about 75 and 125~km, depending on the hub height of the WT.

The RAS observatory will in most cases not be situated in a completely flat terrain. We made a case study using the Effelsberg 100-m telescope, which is situated in the Eifel mountains in the western part of Germany as an example of a telescope in a mountainous region. The dish is located in a valley, which provides a certain level of natural shielding against RFI from all kind of surrounding terrestrial sources, in particular when these are close to the ground. The diffraction on hill and mountain tops along the propagation paths to a WT farm can also substantially attenuate the signal from there. As a result, the required separation distances become much smaller, of the order of 20 to 30~km, depending on the azimuthal direction (because the terrain is not isotropic with respect to the RAS station location) and height of the WT.

\section*{Acknowledgments}

\noindent We thank Alex Kraus for carefully proof-reading our manuscript and Carol Wilson, Dietmar Gaul, Thomas Hasenpusch, and Armin Fleckenstein for fruitful discussions. Furthermore, we would like to express our gratitude to the developers of the many C/C++ and Python libraries, which are made available as open-source software and which we used: most importantly, NumPy \citep{NumPy}, SciPy \citep{SciPy}, Cython \citep{Cython}, and Astropy \citep{Astropy}. Figures were prepared using matplotlib \citep{Matplotlib}. LIDAR-based topography data was kindly provided free-of-charge by the states Northrhine-Westphalia (Land NRW 2018; data available at \url{https://www.opengeodata.nrw.de/produkte/geobasis/dgm/dgm1/}) and Rhineland-Palatinate (GeoBasis-DE / LVermGeoRP 2018; data available at \url{https://lvermgeo.rlp.de/de/geodaten/opendata/}). In both cases, the ``Data license Germany -- attribution -- Version 2.0'' applies (\url{https://www.govdata.de/dl-de/by-2-0}). NASA SRTM topography data was downloaded from \url{http://www.viewfinderpanoramas.org/}, which provides reprocessed tiles with better artifact removal in mountainous environment.

\appendix{Comparison between SRTM and LIDAR-based topography maps}\label{app:srtm_vs_lidar} 

\noindent Many spectrum management compatibility studies make use of terrain height profiles from various sources to calculate the path propagation loss for specific locations \citepalias[e.g.][also this study, see Sec.~\ref{sec:casestudy}]{itu_f1766_0,itu_ra2332}. The topography is of great impact on the path attenuation, especially in hilly or mountainous environments that provide more diffraction edges than flat terrain. A large number of ITU-R recommendations is available that contain methods to estimate the propagation loss in various circumstances (e.g. for different frequency regimes, point-to-point or point-to-area transmission, or for specific use in Monte Carlo simulations); see \citetalias[][and references therein]{itu_p1144_9}. Here, we make use of \citetalias{itu_p452_16}, which is applicable for a large range of frequencies (100~MHz to $\sim$100~GHz) and accounts for the most important propagation channels (see Sec.~\ref{subsec:pathprop}). Nevertheless, even \citetalias{itu_p452_16}, has its shortcomings, as it accounts only for a two-dimensional height profile, but in reality the radio wave could also travel sideways around a mountain by means of diffraction, or be reflected at surfaces left and right of the direct path.

\begin{figure}[!t]
\begin{center}
\includegraphics[width=0.98\textwidth,viewport=25 38 710 644,clip=]{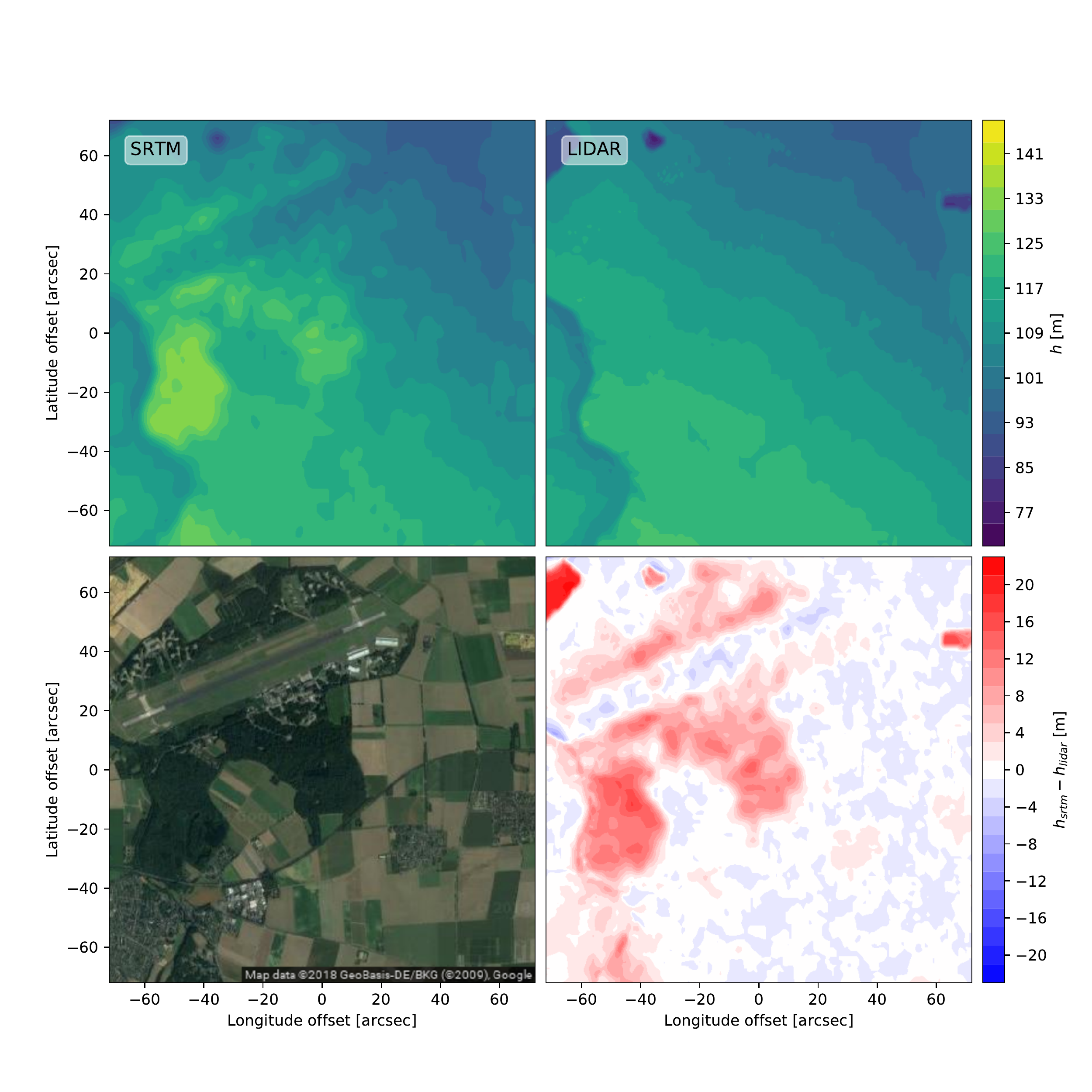}
\end{center}
\caption{Comparison between SRTM and LIDAR-based topography data for a small area around $(l, b) = (6.6681^\circ, 50.8217^\circ)$. The top row contains SRTM and LIDAR terrain heights, the bottom right panel visualizes their difference, and the bottom left panel contains satellite imagery. Local clutter such as forests or villages was not removed from SRTM, as can be seen by comparing the residual image (bottom right) with the satellite image. Note, for NRW, LIDAR-based topography data with a spatial resolution of 1~m is available, but for the comparison with SRTM a lower-resolution version was used ($\sim$ $30~\mathrm{m}$). }%
\label{fig:srtm_lidar_clutter}%
\end{figure}

The topography data recorded by the Shuttle Radar Topography Mission \citep[SRTM][]{farr07} is widely in use today, because of its large area coverage (most of Earth's land mass), good spatial resolution (up to $\sim$ $1"\times1"$, or $30~\mathrm{m}\times30~\mathrm{m}$), and relatively small height errors ($<$ $10~\mathrm{m}$). However, it is well known that SRTM runs into problems in mountainous terrain \citep{berthier06,gupta14,kolecka14}, caused e.g. by high slope angles for the incident RADAR signal. As the Effelsberg 100-m telescope is situated in a valley in the Eifel, we examined the SRTM data close to the site in more detail. For western Germany (North Rhine-Westphalia, NRW, and Rhineland-Palatinate, RP) we are in the fortunate situation that high-resolution topography data from Light Detection and Ranging (LIDAR) observations are available. LIDAR uses laser pulses for range determination, e.g. from an air-plane (which has GPS on-board) and can deliver accurate 3D models of the ground. In conjunction with other auxiliary data sets it was even possible to remove clutter (buildings, trees, etc.) from the topography maps, which was not done for SRTM data. This effect is displayed in Fig.~\ref{fig:srtm_lidar_clutter}, with SRTM and LIDAR-based data in the top left and right panels. The bottom right panel of Fig.~\ref{fig:srtm_lidar_clutter} contains the difference, and the bottom left panel shows satellite imagery. By comparing the two bottom panels, one can see that wood land and housings have left an imprint on the SRTM data.

\begin{figure}[!t]
\begin{center}
\includegraphics[width=0.98\textwidth,viewport=25 38 710 644,clip=]{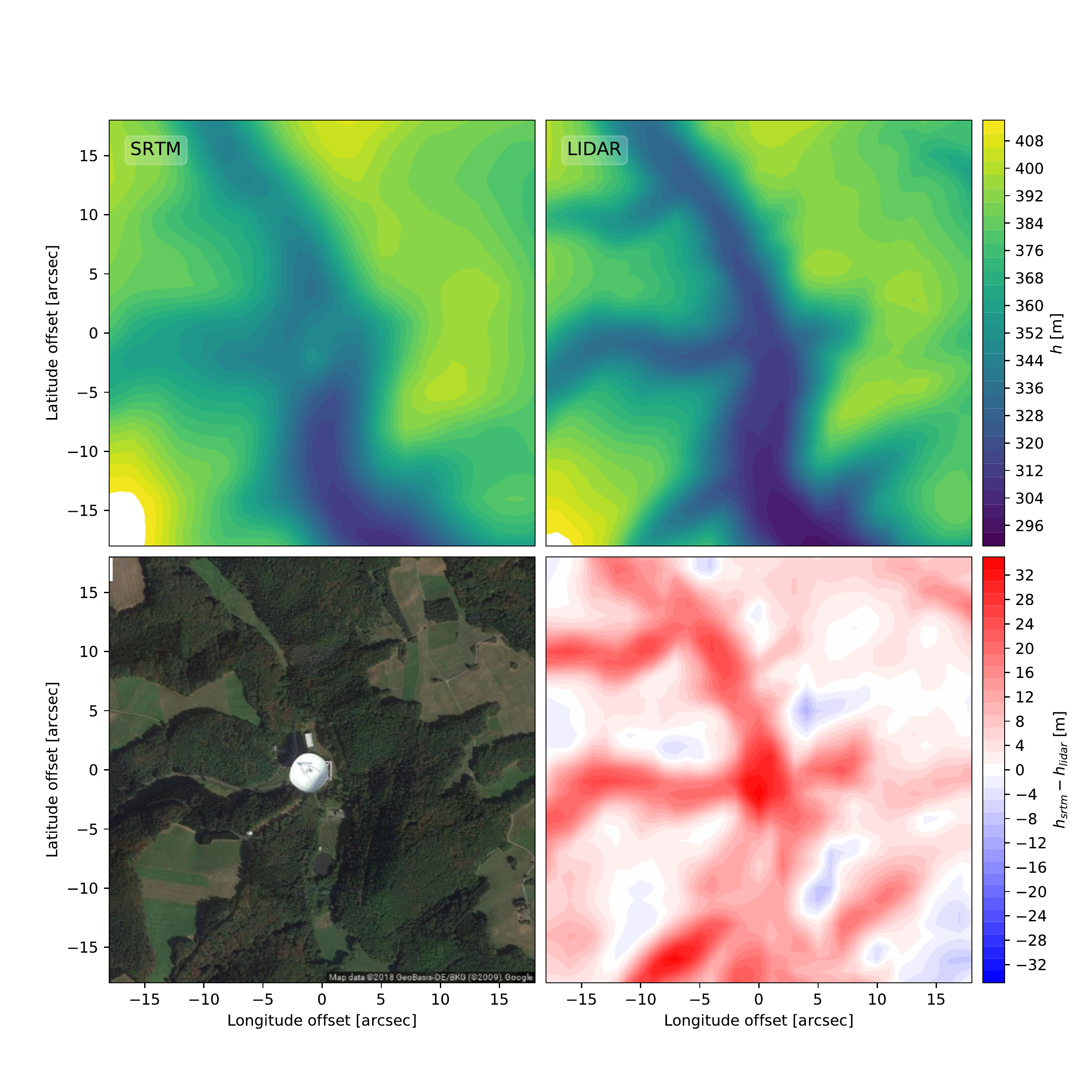}
\end{center}
\caption{As Fig.~\ref{fig:srtm_lidar_clutter} for a region around the Effelsberg 100-m telescope. In SRTM not only the valleys are less deep than for LIDAR-based topography, but the antenna itself has led to a substantial increase ($\sim30- 50~\mathrm{m}$) of the measured terrain height.}%
\label{fig:srtm_lidar_effbg}%
\end{figure}

In Fig.~\ref{fig:srtm_lidar_effbg} the Effelsberg site topography is visualized in the same way as for Fig.~\ref{fig:srtm_lidar_clutter}. Again, SRTM and LIDAR-based topography differ significantly, especially the valleys appear to be more shallow in the former data set. Our comparison also indicates that the Effelsberg 100-m antenna appears to have a significant impact on the SRTM heights, creating an artifact that ought to be removed from the SRTM data before using it for later path propagation calculations. The relative height of the first diffraction edges (adjacent hill tops) would otherwise be effectively smaller which leads to an underestimation of attenuation values. Alternatively, one could also directly work with LIDAR-based topography data, which provide much cleaner and more accurate information --- if such data is available for the area of interest.

\bibliography{references.bib}

\end{document}